\documentclass{mem}
\usepackage{natbib}\usepackage{txfonts}\usepackage{balance}
\usepackage{graphicx}
\usepackage[a4paper]{hyperref}
\idline{75}{282}
\begin{document}
\def\teff{$T\rm_{eff }$}
\def\kms{$\mathrm {km s}^{-1}$}
\def\micron{$\mu\mathrm {m}$}
\def\nev{\mbox{[\ion{Ne}{\sc V}]\,24\,$\mu$m}}

\title{
Unveiling the nature of Seyfert nuclei with 1 - 100 micron spectral
energy distributions
}

   \subtitle{}

\author{
C.\ L. \,Buchanan\inst{1} 
\and J.\ F.\, Gallimore\inst{2}
\and C.\ P.\, O'Dea\inst{3}
\and S.\ A.\, Baum\inst{4}
\and D.\ J.\, Axon\inst{3}
\and A.\, Robinson\inst{3}
\and J.\, Noel-Storr\inst{3}
\and A.\, Yzaguirre\inst{2}
\and M.\, Elitzur\inst{5}
\and M.\, Elvis\inst{6}
\and C.\, Tadhunter\inst{7}
          }

  \offprints{C. Buchanan}

\institute{
University of Melbourne,
Victoria, 3010, Australia
\email{clb@unimelb.edu.au}
\and
Bucknell University,
Lewisburg, PA, 17837, USA
\and
Rochester Institute of Technology,
84 Lomb Memorial Drive, Rochester, NY, 14623, USA 
\and
Rochester Institute of Technology,
54 Lomb Memorial Drive, Rochester, NY, 14623, USA 
\and
University of Kentucky,
Lexington, KY, 40506, USA
\and
Harvard-Smithsonian Center for Astrophysics, 
60 Garden Street, Cambridge, MA, 02138, USA
\and
University of Sheffield,
Sheffield, S3 7RH, UK
}

\authorrunning{Buchanan et al.}

\titlerunning{Seyfert IR SEDs}

\abstract{
The infrared is a key wavelength regime for probing the dusty,
obscured nuclear regions of active galaxies. We present results from
an infrared study of 87 nearby Seyfert galaxies using the Spitzer
Space Telescope and ground-based telescopes.  Combining detailed
modelling of the 3 - 100 micron spectral energy distributions with
mid-IR spectral diagnostics and near-infrared observations, we find
broad support for the unified model of AGNs.  The IR emission of
Seyfert 1s and 2s is consistent with their having the same type of
central engine viewed at a different orientation.  The nature of the
putative torus is becoming clearer; in particular we present evidence
that it is likely a clumpy medium.  Mid-infrared correlations between
tracers of star formation and AGN ionizing luminosity reveal the
starburst-AGN connection implied by the black hole/bulge mass
relation, however it is not yet clear if this is due to feedback.
\keywords{Galaxies: Seyfert -- Galaxies: Active --
Galaxies: Stellar content -- Infrared: galaxies }
}
\maketitle{}

\section{Introduction}

The infrared wavelength regime provides a unique probe into the
nuclear regions of active galaxies (AGNs), as the circumnuclear dust
absorbs and reradiates higher energy photons from the nucleus. We are
conducting a study of the infrared emission from nearby Seyfert
galaxies, using the Spitzer Space Telescope and ground-based
telescopes \citep{buc06}. The goal is to construct matched-aperture
infrared (IR) spectral energy distributions (SEDs) of the central few
kiloparsecs of a large sample of nearby Seyfert galaxies, and to
decompose the SEDs into the contributions to the dust heating from the
active nucleus and any circumnuclear starburst.

\section{Seyfert Infrared SEDs} \label{sec:sey}

\begin{figure*}[t!]
\resizebox{\hsize}{!}{
\includegraphics[clip=true,width=3.7cm]{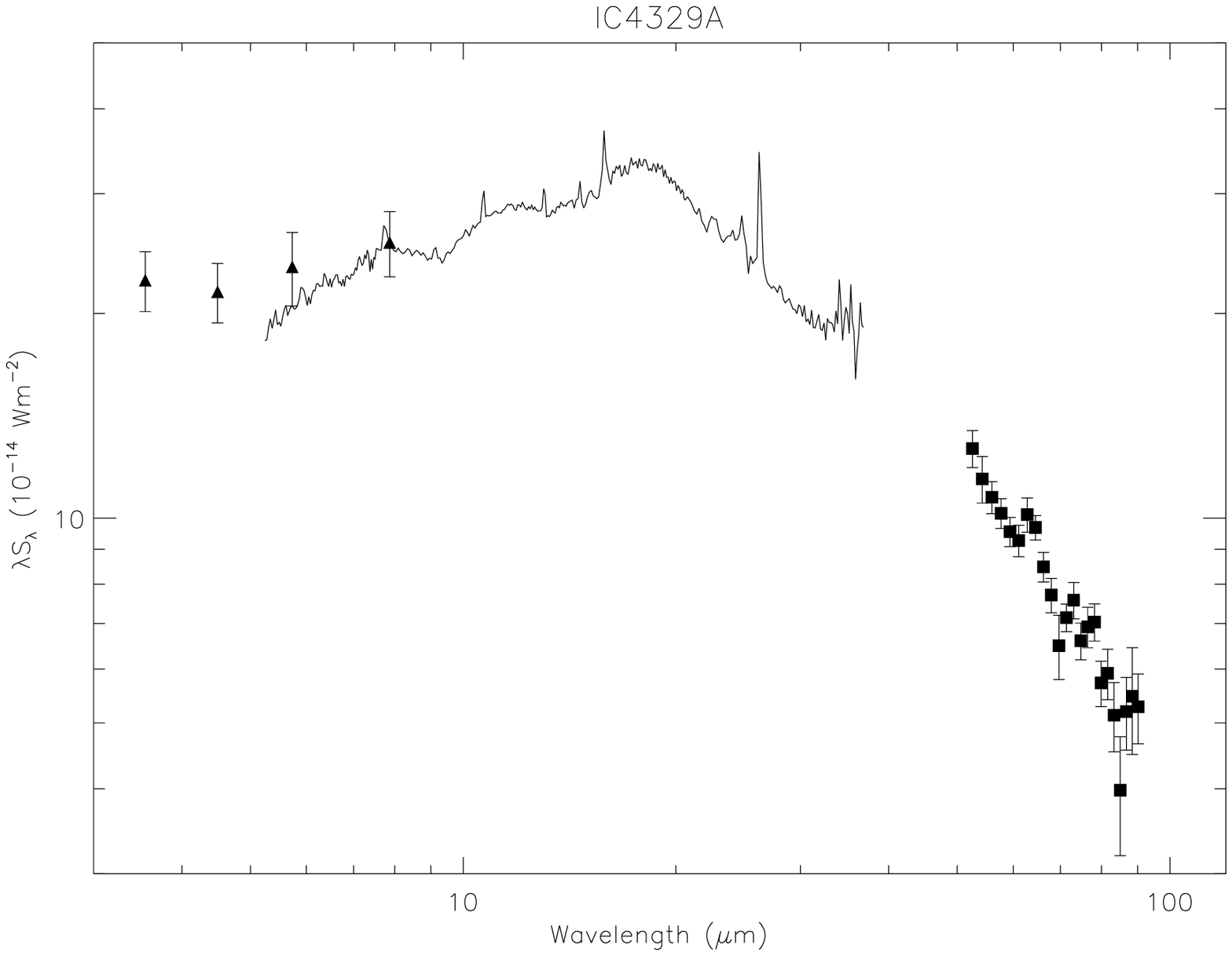}
\includegraphics[clip=true,width=3.7cm]{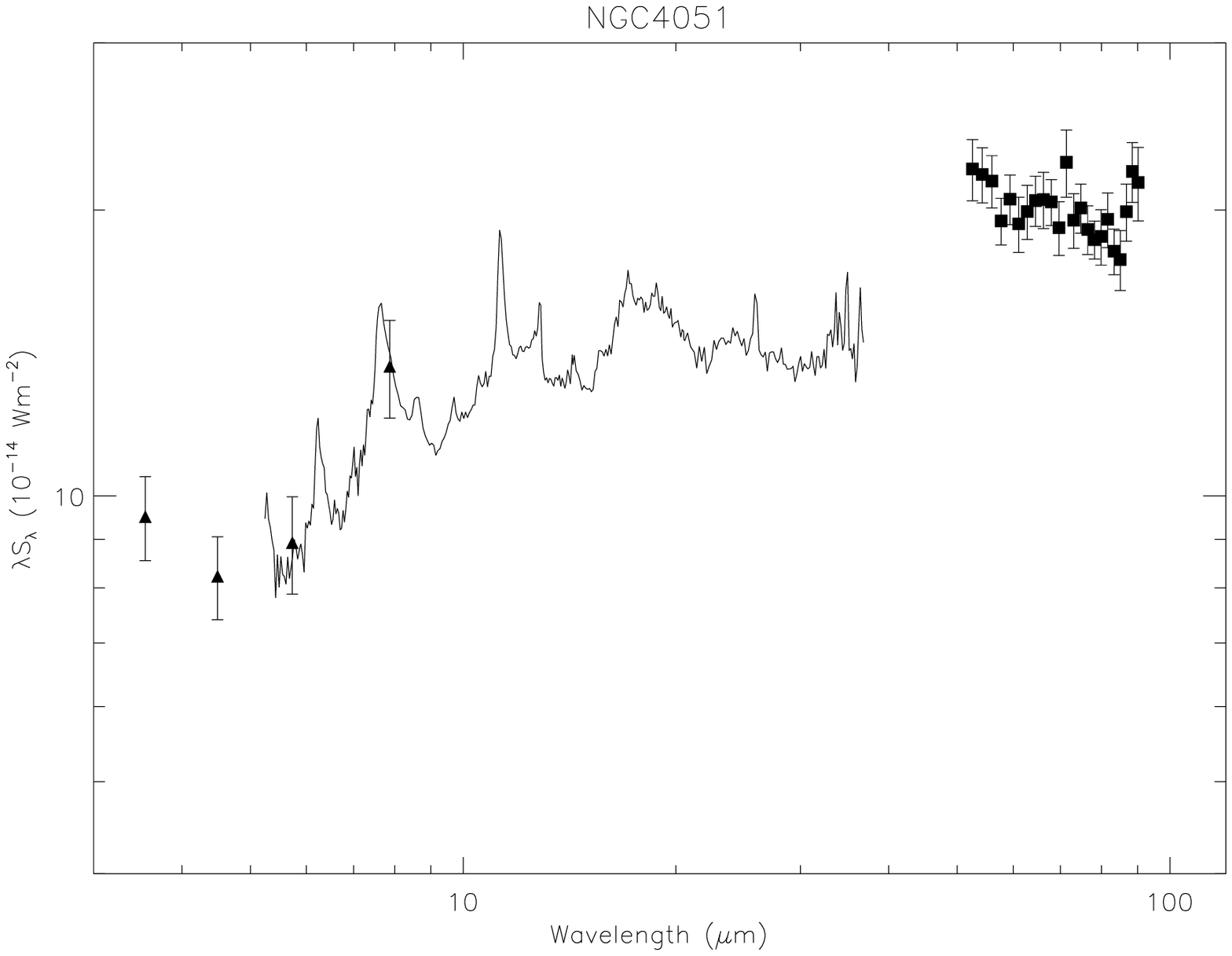}
\includegraphics[clip=true,width=3.7cm]{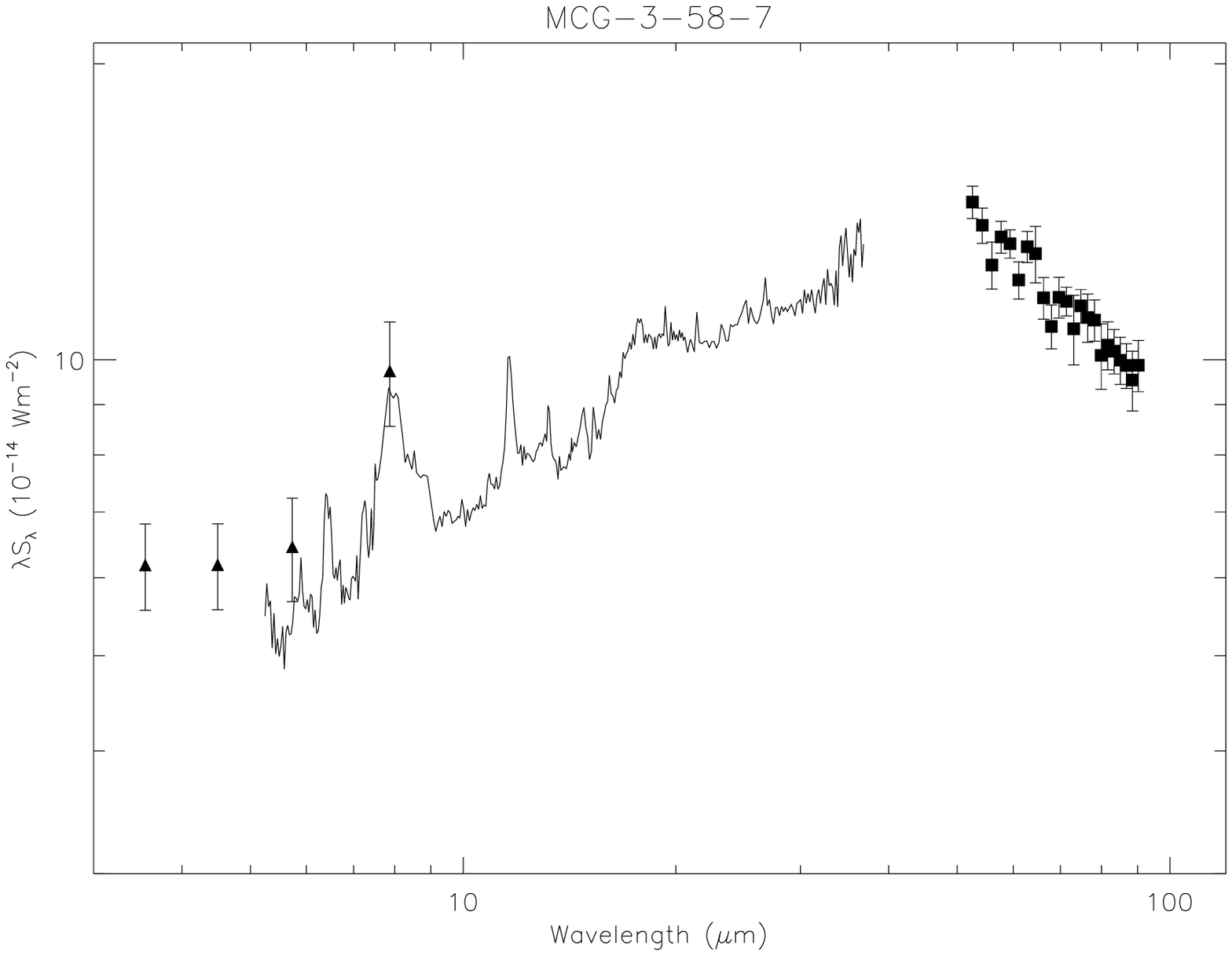}
}
\resizebox{\hsize}{!}{
\includegraphics[clip=true,width=3.7cm]{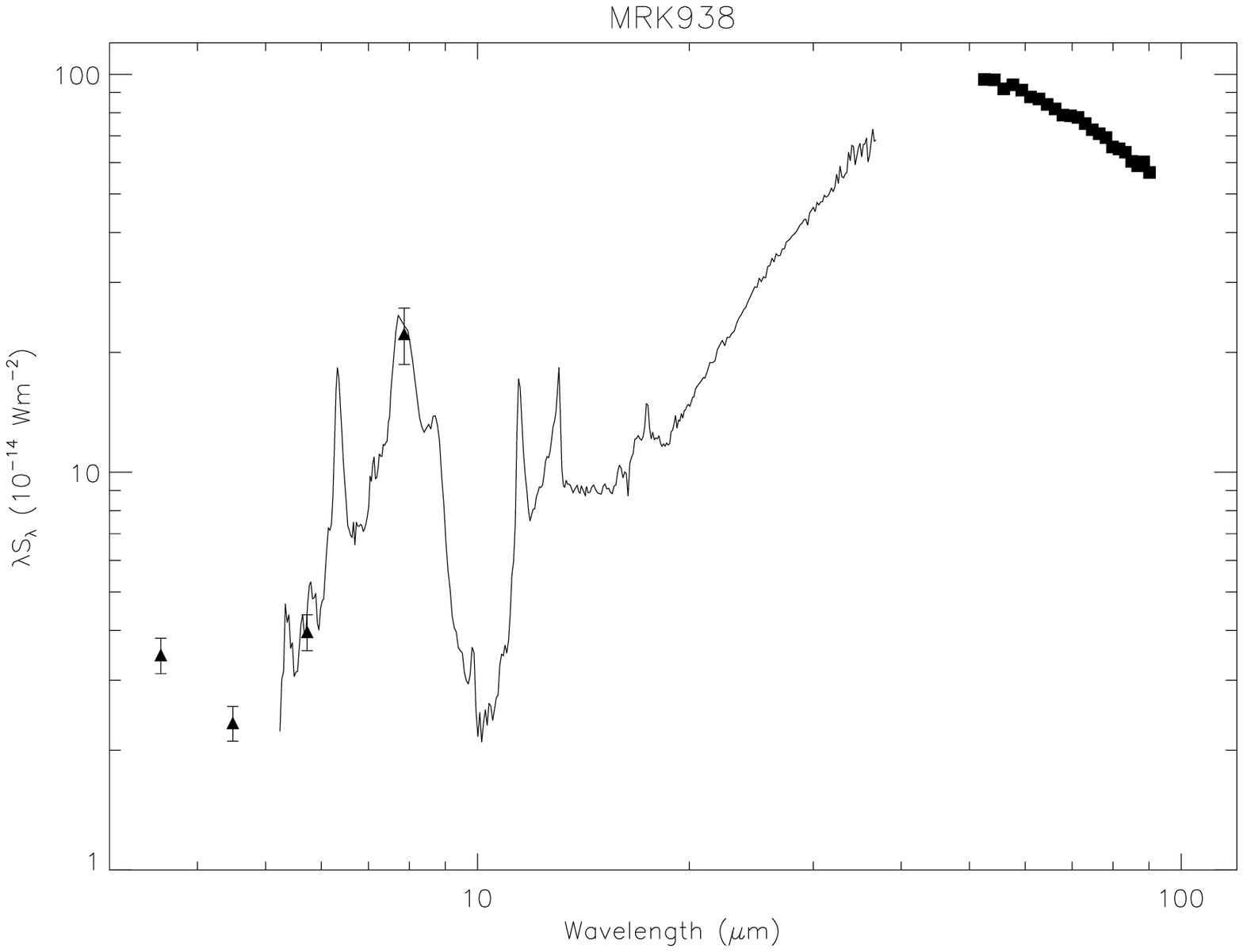}
\includegraphics[clip=true,width=3.7cm]{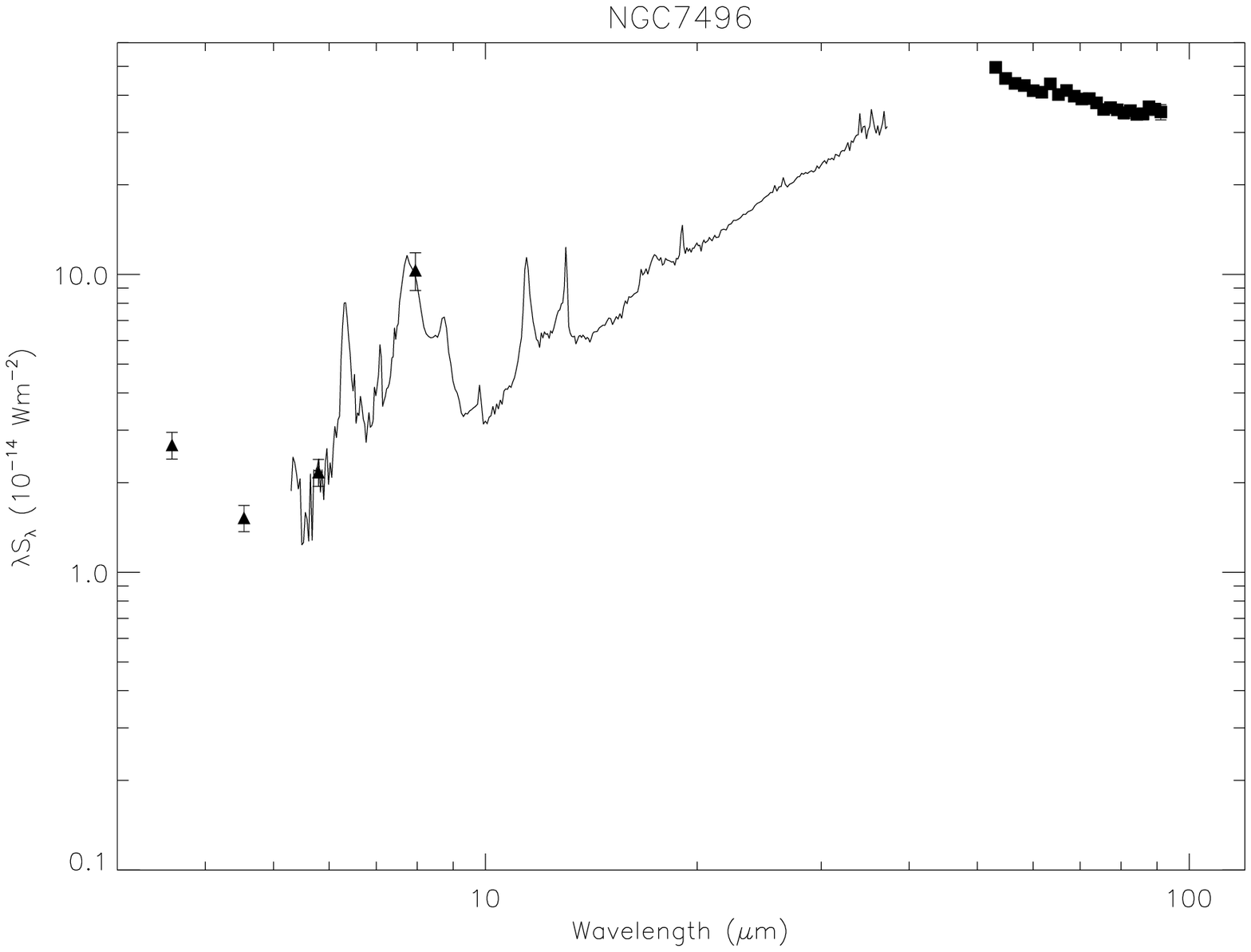}
\includegraphics[clip=true,width=3.7cm]{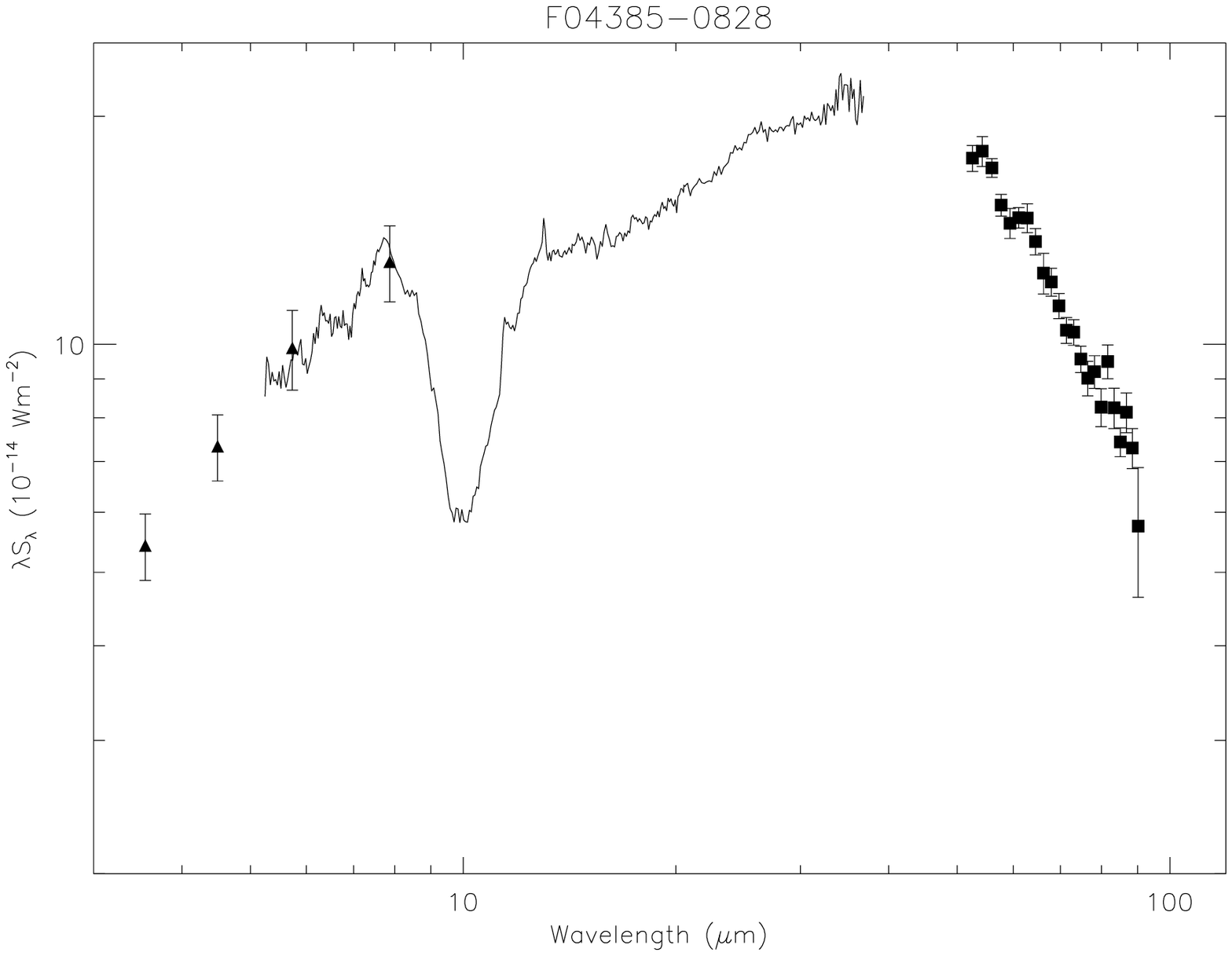}
}
\caption{\footnotesize Spitzer spectral energy distributions of
  several galaxies in the sample, illustrating the common
  features. Silicate features at 9.7 and 18~\micron\ are seen weakly
  in emission (e.g., IC\,4329A). Somewhat stronger silicate absorption
  is also seen, usually in conjunction with PAH emission features
  (e.g., MRK\,938). Very strong silicate absorption (e.g.,
  F04385-0828) is very rare in the sample, and is associated with
  edge-on galaxies, suggesting that some of the absorption is occuring
  in the host galaxy disk. PAH emission features, associated with star
  formation, are very common in the sample (e.g., NGC\,4051,
  MCG\,-3-58-7, MRK\,938, and NGC\,7496). It can been seen that the
  mid- and far-IR spectral indices give an indication of the relative
  contribution of cool dust (peaking in the far-IR) to the SED: bluer
  slopes imply less cool dust emission.  }
\label{fig:seds}
\end{figure*}

The sample of objects studied comprises the 85 Seyfert galaxies from
the extended 12~\micron\ sample \citep{rus93} with $cz < 10000$\kms,
excluding two sources due to problems with the data.  The properties
of the sample are discussed in more detail in \citep{buc06} and Baum
et al.\ (2008, in preparation).
The Spitzer SEDs cover the wavelength range 3.6 -- 100~\micron, and
comprise IRAC photometry at
3.6, 4.5, 5.8 and 8.0~\micron, IRS low-resolution spectroscopy from 5
-- 35~\micron, and MIPS SED-mode spectra from $\sim$50 --
100~\micron. In addition, near-IR photometry measured using UKIRT was
used to construct higher spatial resolution SEDs from $\sim$1 --
4~\micron\ for 40\% of the sample.  Example Spitzer SEDs are shown in
Figure \ref{fig:seds}.

\begin{figure*}[t!]
\resizebox{\hsize}{!}{
\includegraphics[clip=true,width=3.7cm]{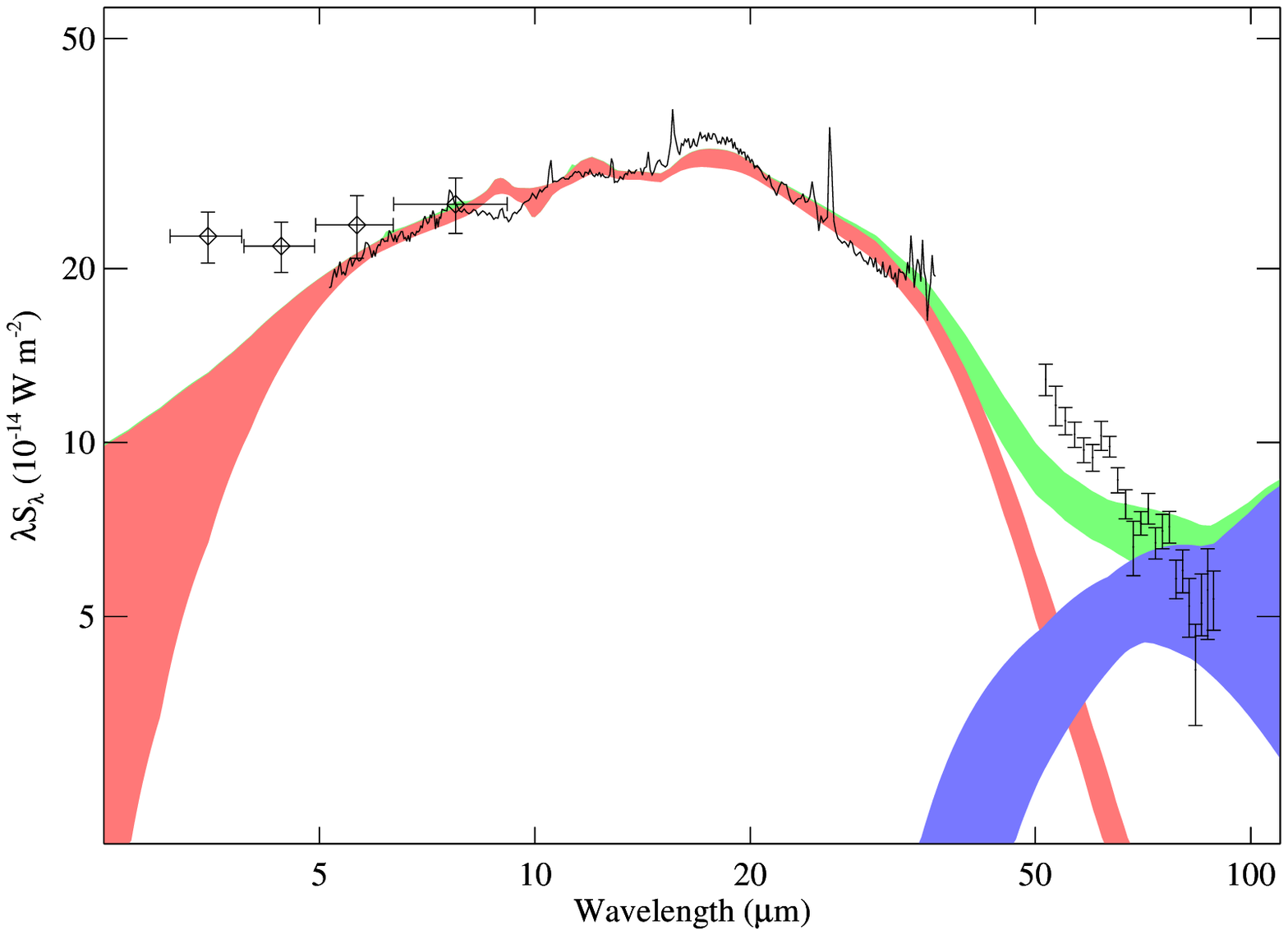}
\includegraphics[clip=true,width=3.7cm]{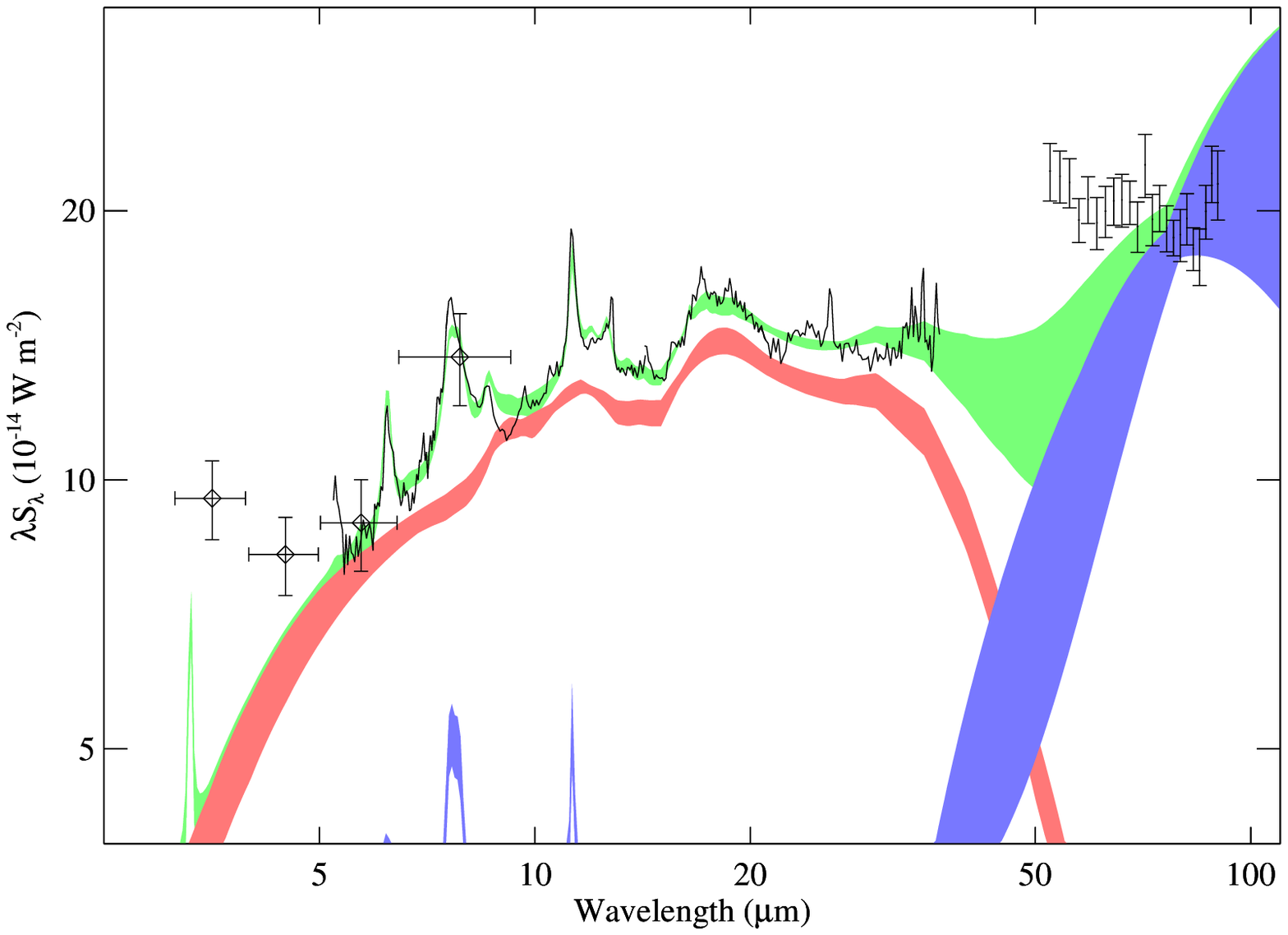}
\includegraphics[clip=true,width=3.7cm]{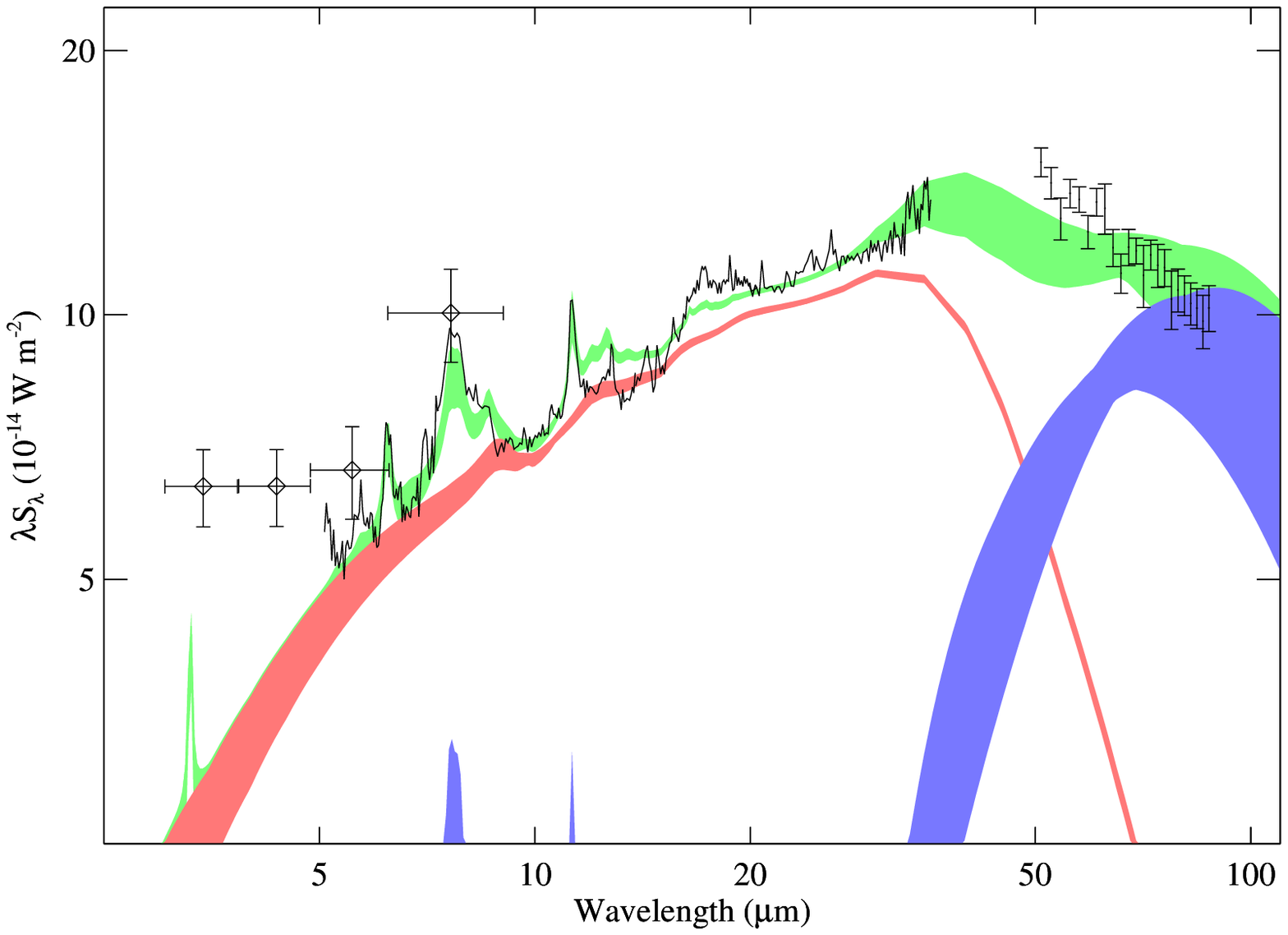}
}
\resizebox{\hsize}{!}{
\includegraphics[clip=true,width=3.7cm]{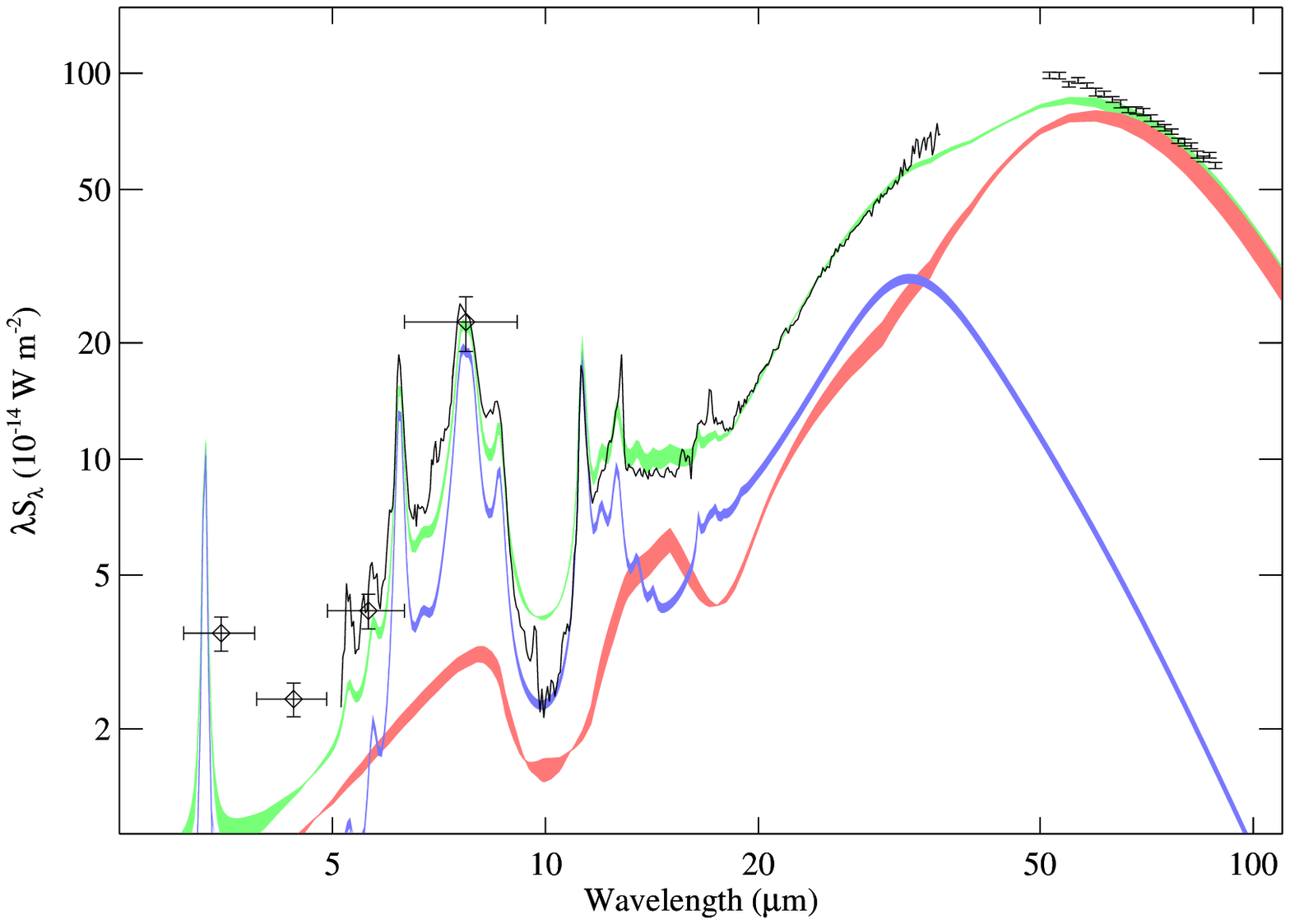}
\includegraphics[clip=true,width=3.7cm]{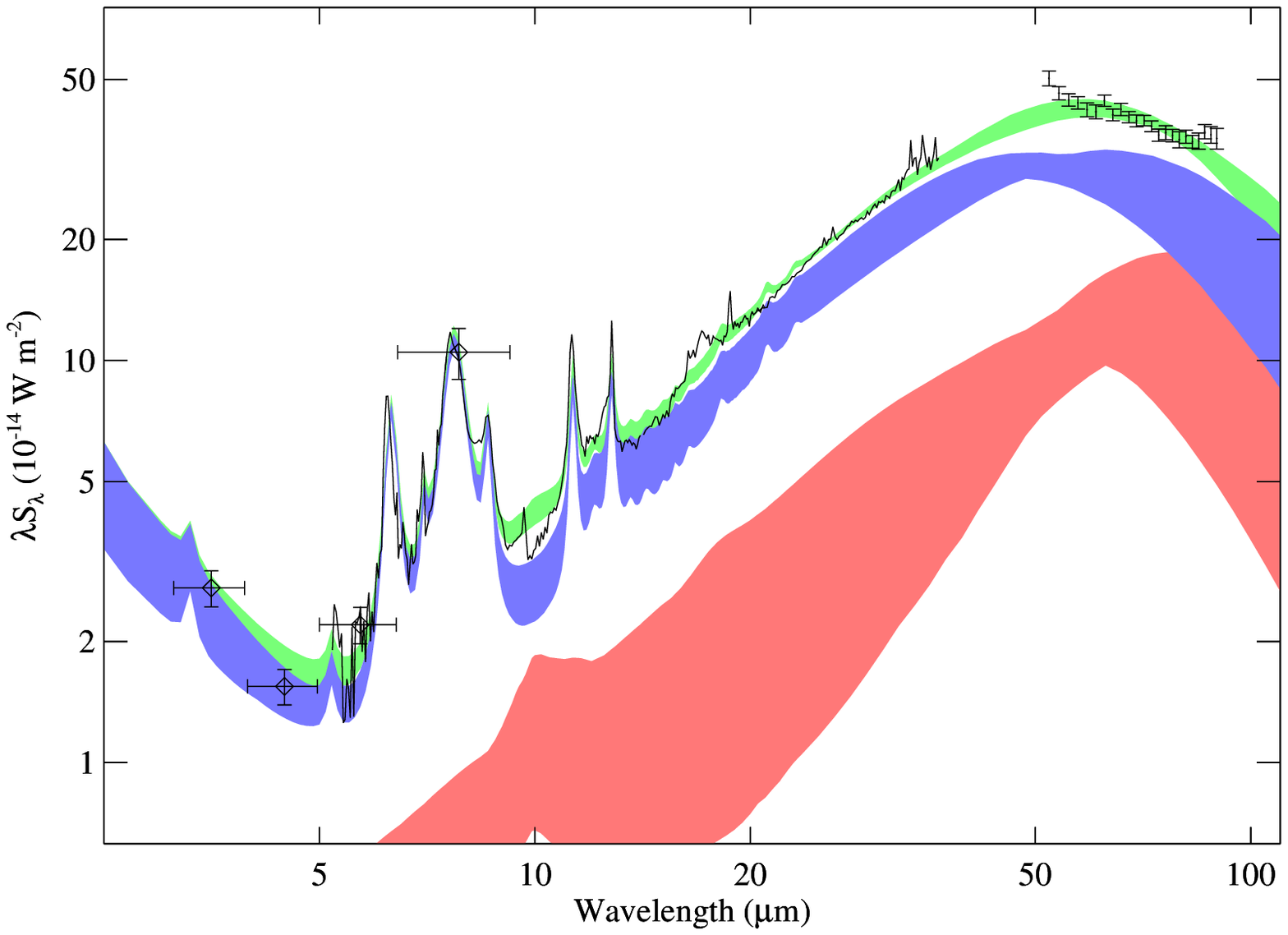}
\includegraphics[clip=true,width=3.7cm]{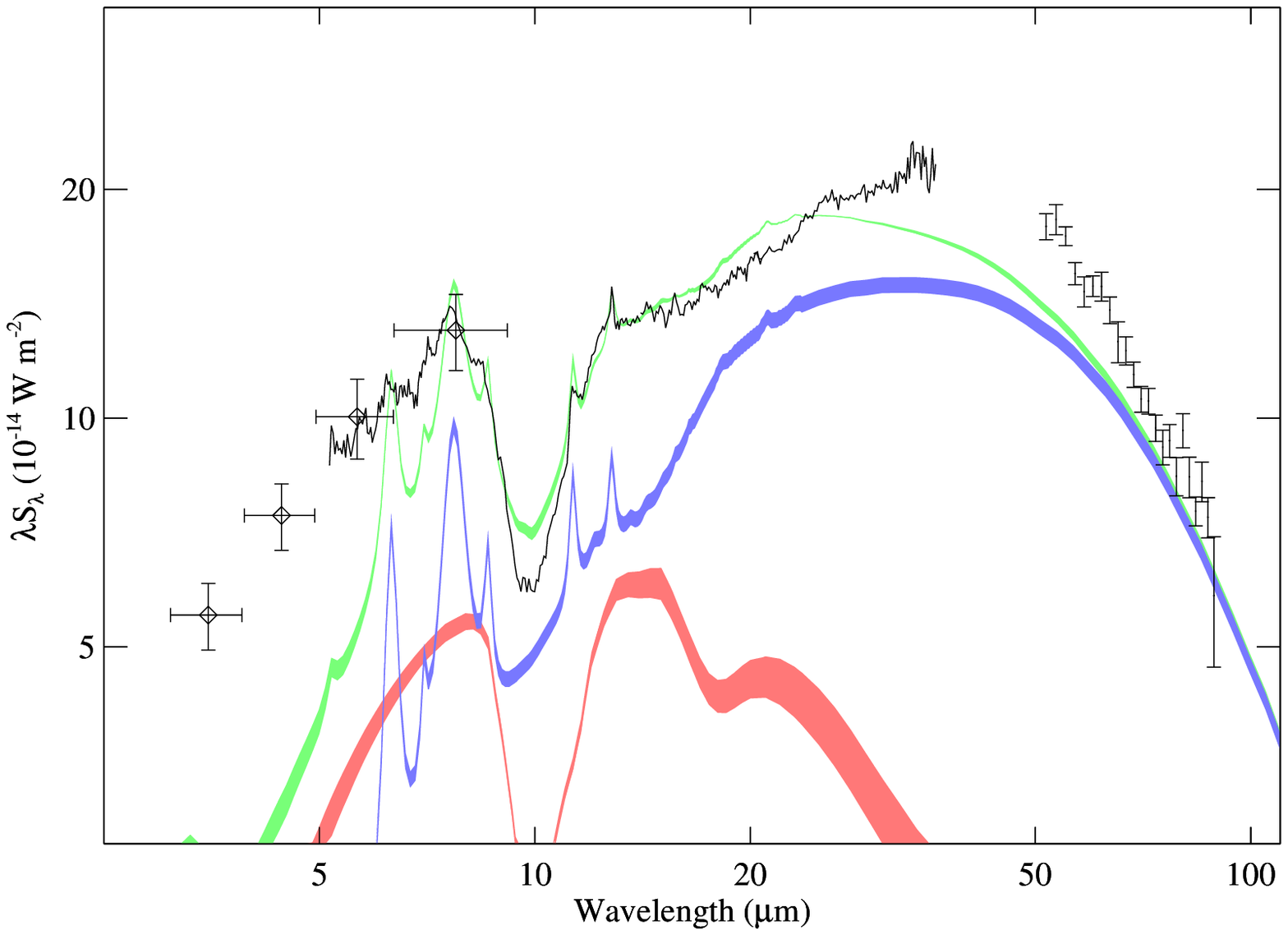}
}
\caption{\footnotesize SED decompositions for the objects whose SEDs
  are shown in Figure \ref{fig:seds}.  The colored bands represent the
  range of SED values afforded by the acceptable fits (at 90\%
  confidence) to the observed SED 
  (black lines and symbols). The range of clumpy torus models is
  shaded red; starburst, blue; and the combination in green.}
\label{fig:mod}
\end{figure*}

The SEDs were decomposed into starburst and AGN contributions to the
dust heating (\citealt{gal07} and Gallimore et al.\ 2008, in
preparation) by fitting the combination of a clumpy torus
\citep{nen02} and a starburst model \citep{sie07}. SED decompositions
for the sources shown in Figure \ref{fig:seds} are shown in Figure
\ref{fig:mod}. 

We find that the Seyfert 1s tend to show weak silicate emission, while
the Seyfert 2s show somewhat stronger silicate in absorption. This is
consistent with the unified scheme for AGN. The sample contains
several `anomalous' sources, including Sy 1s with silicate in
absorption, Sy 2s with silicate in emission and silicate absorption at
9.7~\micron\ but emission at 18~\micron. These features can be
accounted for by the clumpy torus model but not by smooth-density
torus models, providing strong support for the clumpy torus model.

Hidden Seyfert 1s (optically identified Seyfert 2s with a hidden broad
line region (HBLR) observed in the infrared or polarised light) closely
resemble Seyfert 1 galaxies in their infrared SEDs.  This provides
further support for the hypothesis of the unified scheme, that type 1 and 2
AGNs contain the same sort of central engine.

We find that the Seyfert 1s and the Seyfert 2s with known hidden broad
line regions are generally AGN-dominated at 12~\micron, while the
Seyfert 2s without HBLR tend to be starburst-dominated. This is likely to be a
selection effect in the sample due to the mild anisotropy (about a
factor of 2) of the mid-IR emission and the selection of the sample at
12~\micron\ \citep{buc06}.

\begin{figure}[t!]
%\resizebox{\hsize}{!}{
\includegraphics[clip=true,width=5cm]{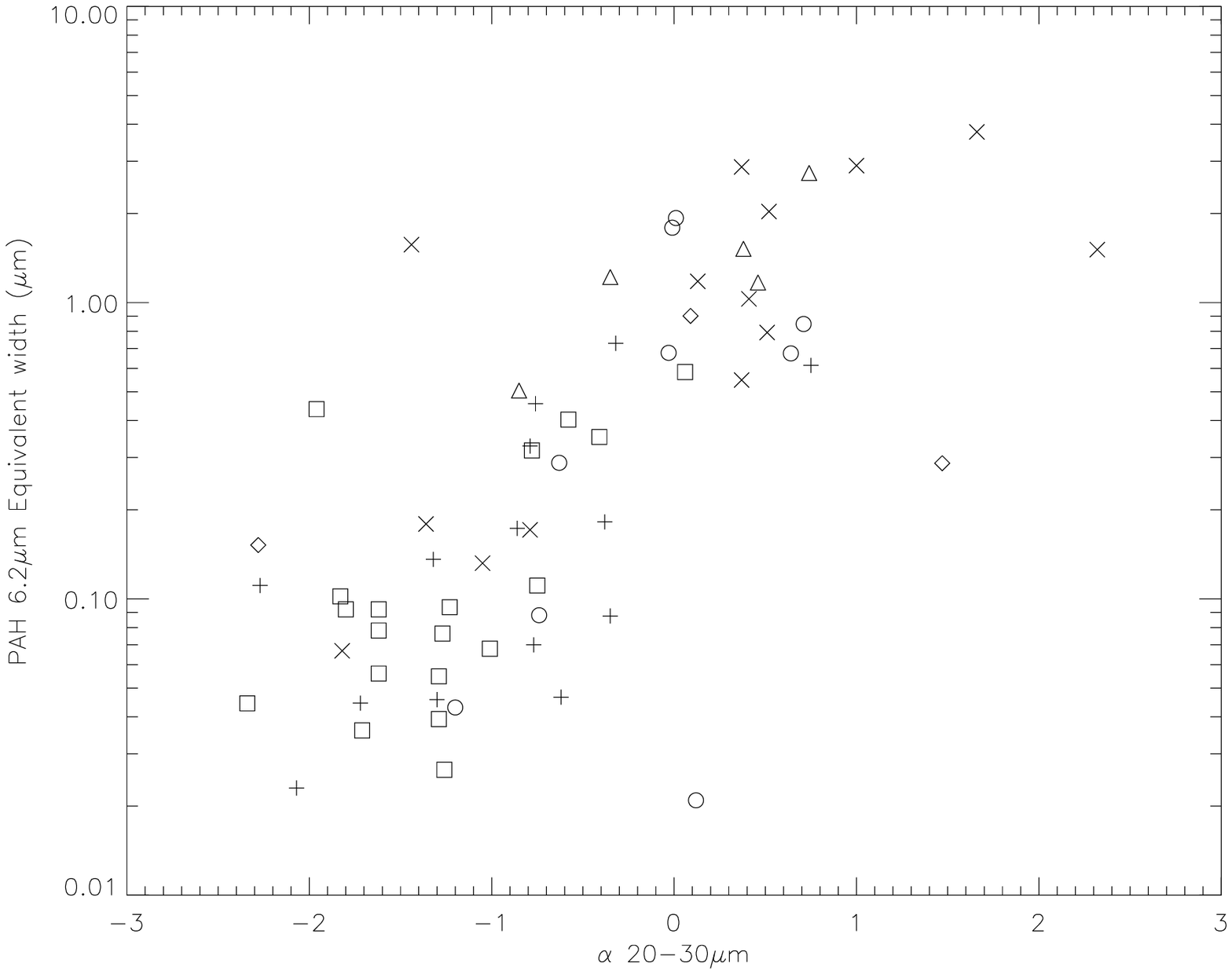}

\includegraphics[clip=true,width=5cm]{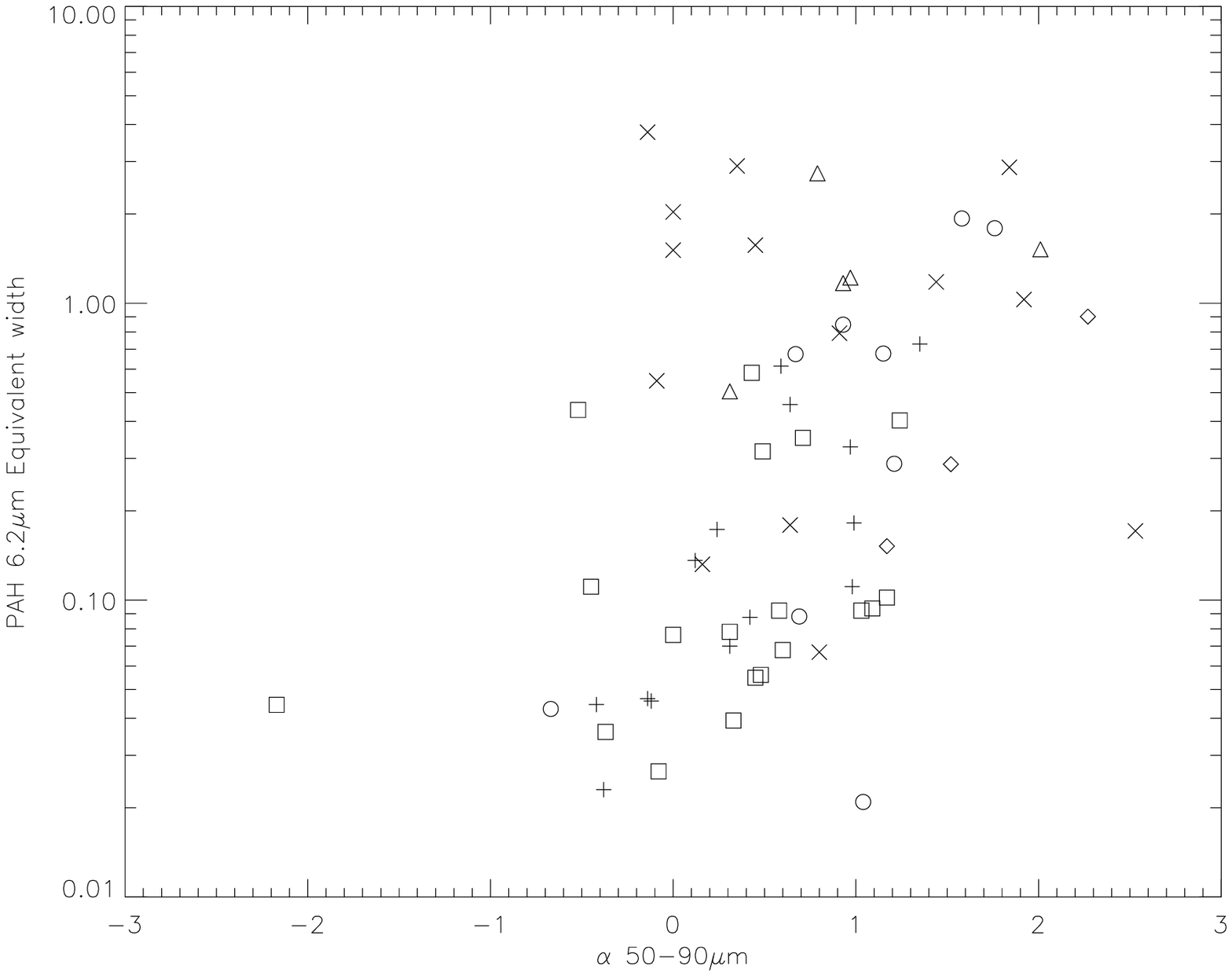}
%}
\caption{\footnotesize
PAH 6.2~\micron\ equivalent width versus 20-30~\micron\ spectral index
{\it (left)} and 50-90~\micron\ spectral index {\it (right)}. The
symbols indicate the Seyfert type: Sy 1, 1.2, 1.5, 1n {\it (squares)};
Sy 1.8, 1.9 {\it (circles)}; Sy 2 with hidden broad line region {\it
  (plus signs)}; Sy 2 without hidden broad line region {\it
  (crosses)}; LINERs {\it (diamond)}; and starburst galaxies {\it
  (triangles)}.
}
\label{fig:alpha}
\end{figure}

\begin{figure}[t!]
%\resizebox{\hsize}{!}{
\includegraphics[clip=true,width=5cm]{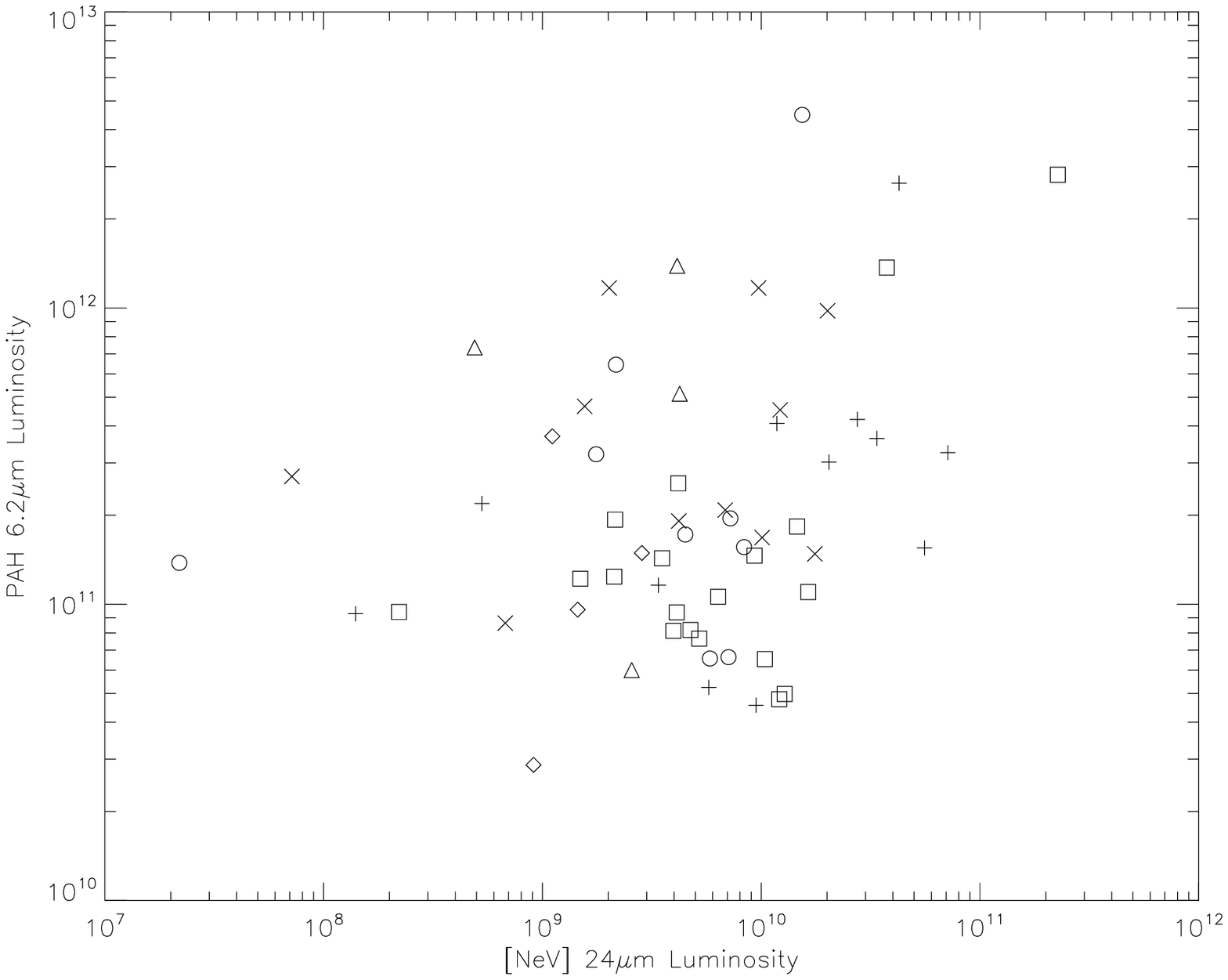}
%}
\caption{\footnotesize
PAH 6.2~\micron\ luminosity vs \protect\nev\ luminosity. The symbols
are the same as for Figure \protect\ref{fig:alpha}.
}
\label{fig:lum}
\end{figure}

Spectral indices in the mid-IR (20-30~\micron) and far-IR
(50-90~\micron) were measured, to estimate the relative contributions
of cool dust.  In addition, line and continuum diagnostics were
measured using PAHFIT \citep{smi07}.  Figure \ref{fig:alpha} shows
that the PAH 6.2~\micron\ equivalent width is correlated with with the
mid- and far-IR spectral indices.  The fact that the PAH emission is
stronger in sources with a larger cool dust contribution (redder
spectral indices) argues that the cool dust is primarily heated by
star formation and not the active nucleus.  We also note that the Sy
2s {\it with} hidden broad line regions show weaker star formation
(weaker PAH emission) than those {\it without} hidden broad line
regions, suggesting that the broad line region is
more difficult to detect in the presence of more star formation.
Correlations between AGN and starburst tracers have been found for
QSOs and Seyferts (e.g., \citealt{haa03,mai07}).  Figure \ref{fig:lum}
shows the observed correlation between the 6.2~\micron\ PAH luminosity
and the \nev\ luminosity for our sample. It is
not clear if this apparent correlation is due to feedback or is simply that
the both properties relate directly to the available supply of
molecular gas.

\begin{figure}[t!]
%\resizebox{\hsize}{!}{
\includegraphics[clip=true,width=5cm]{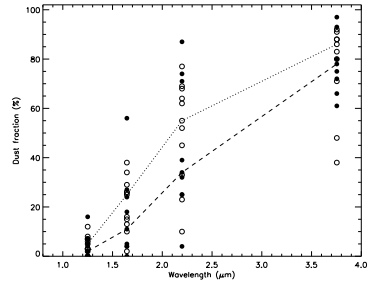}
%}
\caption{\footnotesize
Fraction of the emission contributed by the hot dust (AGN) component,
at the four near-IR wavelengths observed with UKIRT. Filled symbols
indicate Sy 2s and open symbols Sy 1s. The median dust fractions for
type 1s and 2s are shown by the dotted and dashed lines, respectively.
}
\label{fig:nirsed}
\end{figure}

The near-IR SEDs were decomposed into stellar and hot dust components
using a simple non-AGN + blackbody model, where the non-AGN SED was
derived from the colours of non-Seyfert galaxies in the
12~\micron\ sample of galaxies \citep{spi95} and the temperature of
the blackbody was allowed to vary. We find that the dust temperatures were
typically $\sim$1000~K, somewhat cooler than the dust sublimation
temperature. Figure \ref{fig:nirsed} shows the fractional contribution
to the emission by the hot dust component as a function of wavelength.
We find that the AGN contribution increases with wavelength and
dominates by 3.7~\micron.  The Sy 1s show slightly larger hot dust
contribution than the Sy 2s, as expected in the unified scheme.

\section{Conclusions}

We have constructed 20\arcsec\ IR SEDs of a sample of 85 nearby
Seyfert galaxies.  Decomposition of the SEDs into AGN and starburst
contributions provides support for clumpy torus models.  The SEDs are
consistent with the unified scheme for AGN.  We find that the Seyfert
1s are AGN-dominated at 12~\micron\ while the Seyfert 2s are
starburst-dominated, which is likely to be a selection effect. We find
evidence that the cool dust is primarily heated by star formation in
these Seyferts.

\begin{acknowledgements}
Thanks to the organisers for a fantastic conference.  This work is
based in part on observations made with the Spitzer Space Telescope,
which is operated by the Jet Propulsion Laboratory, California
Institute of Technology under a contract with NASA. Support for this
work was provided by NASA through an award issued by JPL/Caltech.  The
United Kingdom Infrared Telescope is operated by the Joint Astronomy
Centre on behalf of the Science and Technology Facilities Council of
the U.K.

\end{acknowledgements}

\bibliographystyle{aa}

\end{document}